\documentclass[onecolumn,nobibnotes]{revtex4}

\usepackage{amsmath}
\usepackage{graphicx}
\usepackage{float}
\usepackage{subfigure}
\usepackage{wrapfig}
\usepackage{multirow}
\usepackage{tabularx}
\usepackage{array}
\usepackage{url}
\usepackage{tikz}
\usetikzlibrary{positioning}

\usepackage{hyperref}
\hypersetup{
    colorlinks=true,
    linkcolor=blue,
    citecolor=blue,
    }

\def\lsim{\raise0.3ex\hbox{$<$\kern-0.75em\raise-1.1ex\hbox{$\sim$}}}
\def\gsim{\raise0.3ex\hbox{$>$\kern-0.75em\raise-1.1ex\hbox{$\sim$}}}

\def\beq{\begin{equation}}
\def\eeq{\end{equation}}
\def\beqa{\begin{eqnarray}}
\def\eeqa{\end{eqnarray}}

\def\gappeq{\mathrel{\rlap {\raise.5ex\hbox{$>$}}
{\lower.5ex\hbox{$\sim$}}}}

\def\lappeq{\mathrel{\rlap{\raise.5ex\hbox{$<$}}
{\lower.5ex\hbox{$\sim$}}}}

\def\Toprel#1\over#2{\mathrel{\mathop{#2}\limits^{#1}}}

\begin{document}

\title{Charmonium production in high multiplicity pp collisions 
and the structure of the proton}

\author{R. Terra$^1$ and F. S. Navarra$^1$}
\affiliation{$^1$Instituto de F\'{\i}sica, Universidade de S\~{a}o Paulo, 
 Rua do Mat\~ao, 1371, CEP 05508-090,  S\~{a}o Paulo, SP, Brazil\\
}
\begin{abstract}
In this work we study charmonium production in high multiplicity proton-proton   
collisions.  We investigate the role of the spatial distribution of partons  
in the protons and assume that the proton has a Y shape. In this configuration  
quarks are more at the surface and gluons in the inner part of the proton.    
Going from peripheral to more central and then to ultra-central proton-proton  
collisions, we go from quark-quark collisions to gluon-gluon collisions. Since  
gluons are much more abundant, the cross sections grow. In the case of charm 
production this growth is enhanced by the fact that, 
$\sigma( g + g \to c + \bar{c}) >> \sigma( q + \bar{q} \to c + \bar{c})$.
These effects can explain the growth seen in the data.
\end{abstract}
\maketitle

\section{Introduction}

Surprising new features of proton proton collisions were revealed when 
the LHC collaborations became able to trigger on very high multiplicity events 
\cite{alice-psi7,alice-cc,star18,alice19,alice20,alice-psi13,alice22}. In these  
events, the data presented evidence of collective behavior, which could be 
interpreted in terms of a hydrodynamical expansion of the system. However,  
to the best of our knowledge, so far no particle production model, 
hydrodynamic or non-hydrodynamic, matches all the  features of the 
high-multiplicity pp data. 

High multiplicity events come from ultracentral collisions, with very small  
impact parameter. In this regime we expect to observe several effects which 
can be responsible for the anomalous features of particle production, such as 
double parton scattering and parton saturation, caused by the larger saturation 
scale at lower impact parameters. Apart from these changes in the dynamics of
the collisions, it is also possible that geometric effects associated with 
the spatial distribution of strongly interacting matter play a significant role. 

Successfull models of the static proton are based on lattice QCD   
simulations, which show that quarks are bound by gluonic strings. This leads 
to the ``Y'' picture of the proton, where quarks are at the extremities tied 
by an  Y-like gluon string, called gluon junction \cite{suga15} or baryon  
junction. This structure leads to a spatial configuration where the gluons 
are mostly in the center and the quark closer to the proton surface.  

At first sight the static geometric configuration of the proton could be
blurred or even completely washed out by a boost  with the consequent parton
evolution and branching. However, there are indications that the geometric
organization of matter persists even at LHC energies. In \cite{manti23}
it was shown that exclusive vector meson production is sensitive to the
geometric deformation of the target. The authors conclude that the fluctuations
in the nuclear geometry originating from the deformed structure are not
washed out by the JIMWLK evolution and hence the deformations previously
inferred from low-energy experiments will be visible in high-energy collisions.
This suggests the the Y shape of the proton could survive the quantum evolution
and manifest itself in high energy proton proton collisions.

The existence of the Y-shape 
configuration of the proton (along with its gluon junction) might shift the 
leading baryon distribution to smaller rapidities, providing an additional 
mechanism of baryon stopping. In these events the baryon number would be 
carried by the baryon junction, whereas in the traditional approach it is 
carried by the valence quarks (for a discussion see \cite{marb89}).  
This interesting idea was proposed in 
\cite{khar96}, implemented in the Monte-Carlo event generator HIJING/B  
\cite{top04} and also in analytical models such as in \cite{bopp06}, 
being successfull in explaining the data on forward baryon production. 
The search for new effects of the baryon junction is in progress \cite{bran22}. 

Having in mind what was said above, we would expect to see more manifestations 
of the proton Y shape. This was explored in ~\cite{deb20}. One   
of the surprising aspects of high multiplicity proton proton collisions is the 
observation of the ridge effect and of elliptic flow. In Ref.~\cite{deb20} a  
simple model based on the proton spatial configuration was developed to      
explain these phenomena.  The authors improved the existing partonic Glauber 
model for proton-proton collisions \cite{loi16}, including anisotropic and 
inhomogeneous density  profiles  for the proton. 
They obtained a very good description of the $v_2$  measured by the ALICE 
collaboration in pp collisions at $\sqrt{s} = 13$ TeV \cite{alice19}.

Another interesting observation made in high multiplicity events refers to 
charm production. In \cite{alice-cc} the ALICE collaboration measured 
the $D^0$ yield as a function of the central rapidity density and found out 
an unexpectedly strong growth with the charged particle multiplicity. 
A similar trend was observed in the case of $J/\psi$ production. There are 
already some possible explanations of this  growth of the charm yield given 
in Refs. \cite{cgc1},  \cite{boris20}, \cite{siddi20} and \cite{tripa22}.  

In this work we will try to understand the data on charmonium production in  
high multiplicity events using the same geometrical model proposed in      
\cite{deb20} with the parameters fixed in that work. Charmonium production 
will be computed with the Color Evaporation Model (CEM). 
Leaving aside the quantitative aspects,
the idea is fairly simple and it is as follows. The proton-proton collisions 
can be described by  Y-Y collisions. Since we are not interested in the 
anisotropy aspects of the collision, we can take the average configuration of 
the Y over different orientations, i.e. we ``rotate'' the Y, obtaining a circular
configuration with an inner gluonic shell and an outer quark shell. In other 
words, we have a core (gluon) - corona (quark) model of the proton. Going from   
peripheral to more central and then to ultra-central, we go from quark-quark   
collisions to gluon-gluon collisions. Since gluons are much more abundant, and 
since $ \sigma (g g \to c \bar{c}) \, >> \, \sigma (q \bar{q} \to c \bar{c})$  
the cross sections grow strongly. These effects combined should explain the 
growth seen in the data. 

In the next section we make some remarks concerning the proton structure, 
in Section III we review  the version of the Glauber model adapted for 
proton-proton collisions and calculate the basic quantities, which are 
$N_{part}$ and $N_{coll}$. In Section IV we review the main formulas 
of the Color Evaporation Model used to study charm production. In 
Section V we show the results and compare them with data. Finally, some 
concluding remarks are presented in the last section. 

\section{Remarks on the proton structure}

\begin{figure}[h!]
\begin{tabular}{ccc}
    \includegraphics[scale=0.4]{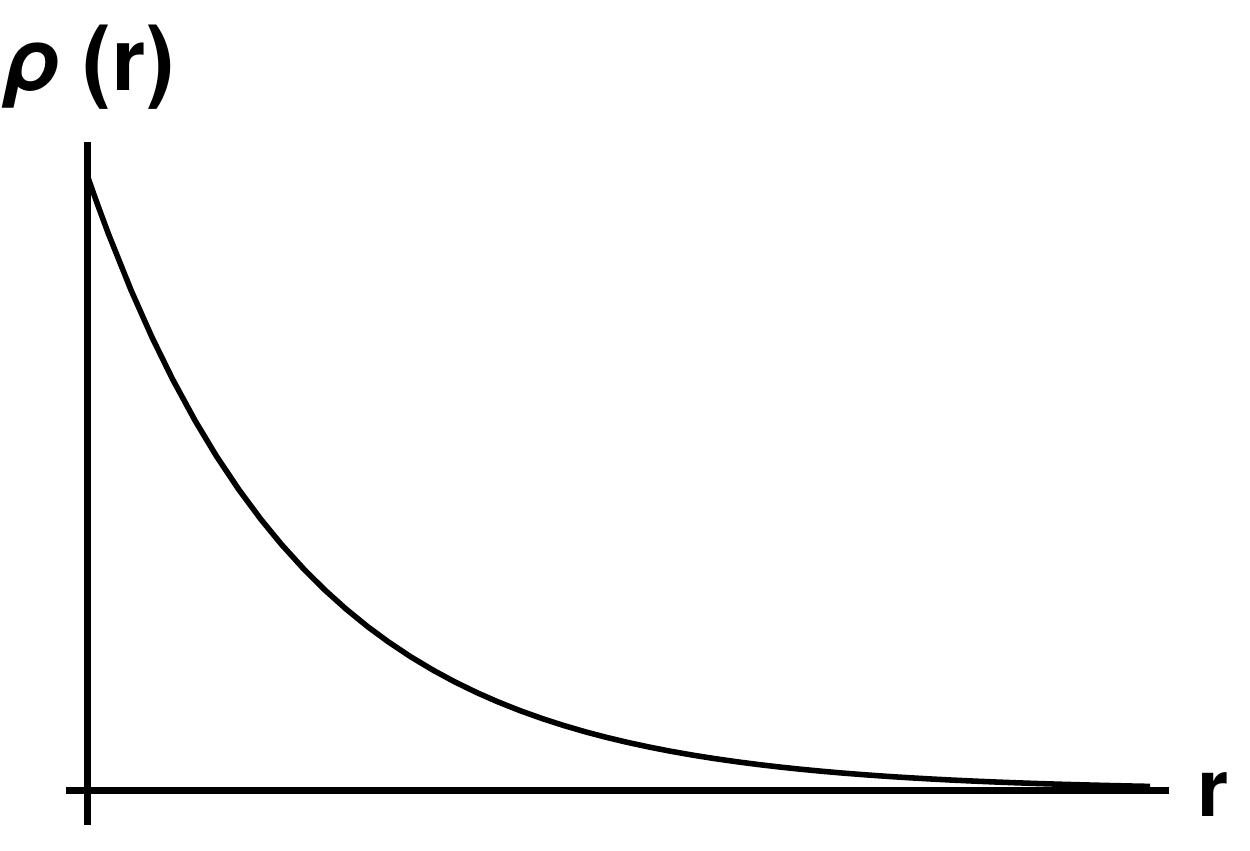}& 
\,\,\,\,\,\,\,\,\,\,\,\,\,\,\,\,\,\,\,\,\,\,\,\, 
\,\,\,\,\,\,\,\,\,\,\,\,\,\,\,\,\,\,\,\,\,\,\,\,
                                              & 
    \includegraphics[scale=0.4]{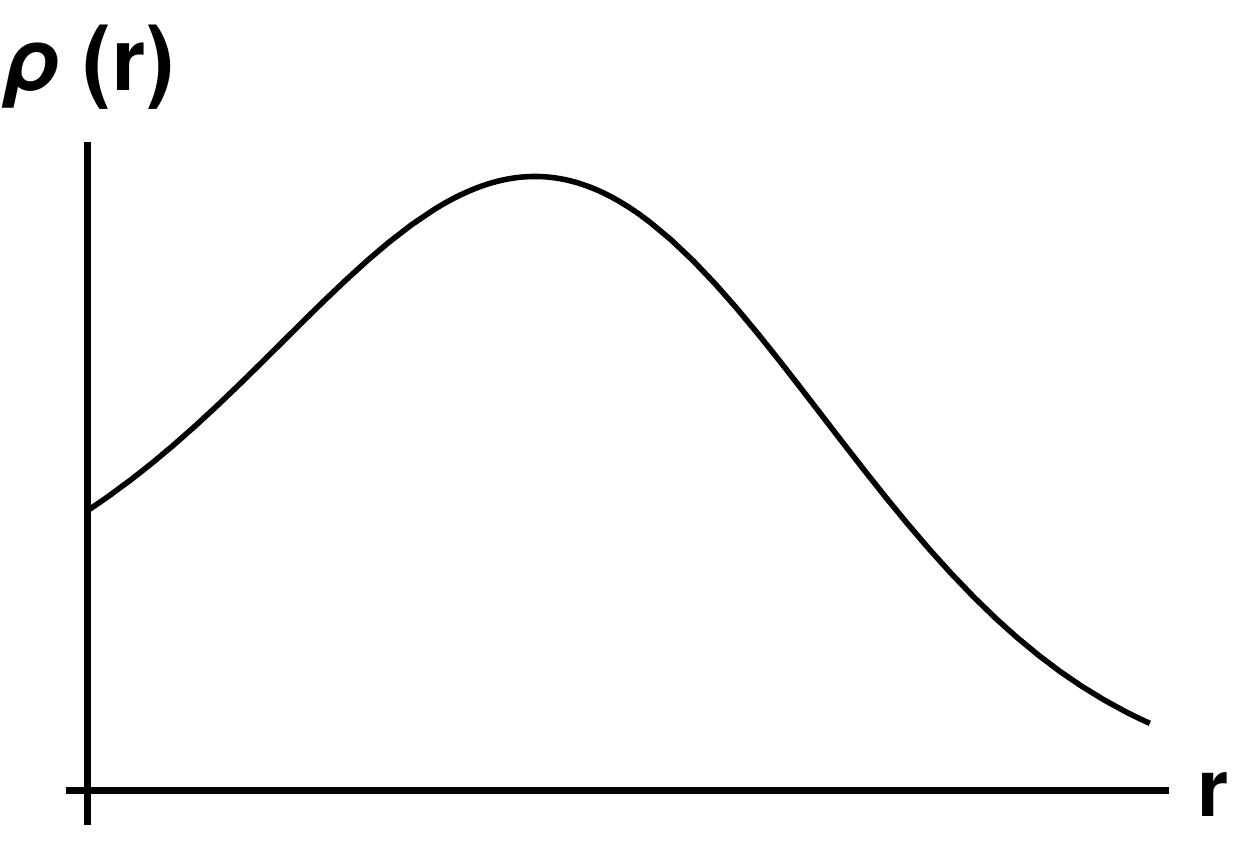} \\
  (a) & \,\,\, & (b)
\end{tabular}
    \caption{a) Charge distribution of the proton measured in elastic 
electron proton scattering at low energies and $Q^2 \simeq 0$. 
b) Charge distribution suggested by the Y shape model of the proton.}
\label{rhocha}
\end{figure}

In textbooks \cite{thomson} (Chap. 7) we learn that in low energy 
elastic $e - p$ 
scattering we can determine the charge distribution of the proton. This 
is done by introducing electric, $G_E(Q^2)$, and magnetic, $G_M(Q^2)$, form 
factors and fitting the resulting (Rosenbluth) formula to the experimental 
data. 
Then, in a very specific limit, when $ Q^2 << m^2_p$, we find that 
$Q^2 \simeq \mathbf{q}^2$ and hence $G_E(Q^2) \simeq  G_E(\mathbf{q}^2)$. In this
limit the electric form factor can be interpreted as the Fourier transform of the
charge  distribution: 
\begin{equation}
G_E (\mathbf{{q^2}}) =
\int \rho(\mathbf{r}) e^{i \mathbf{q} \cdot \mathbf{r}} d^3\mathbf{r}
    \label{ge}
\end{equation}
Hence, measuring $G_E(Q^2)$ we find the $\rho$ shown in Fig.~\ref{rhocha}a. 
Moving away from the very low $Q^2$ limit, we do not have a well defined 
prescription to extract charge densities from the measured form factors 
(see, however, \cite{lorce}  for progress in this direction). From  
electron-proton high energy scattering in the high $Q^2$ limit, we know 
\cite{thomson} (Chap. 8) that the incoming probe (photon or gluon) will indentify
pointlike particles (partons) inside the proton, but we do not know how these 
partons are distributed in the transverse plane. 

In the parton model description of deep 
inelastic scattering, the parton distribution functions depend on $Q^2$ and this
dependence comes from the solution of the DGLAP evolution equations.  
In \cite{gla11,gla12,gla17} a different kind of evolution was proposed: 
the Renormalization Group Procedure for Effective Particles (REGPEP). In this 
approach, increasing the resolution scale we go from a configuration where the  
quarks are unresolved to a configuration with three effective quarks 
(quarks + antiquarks + gluons) disposed in the vertices of an equilateral 
triangle with a star-like juntion between them.  Hence at high $Q^2$ the charge
density tends to be moved from the origin, as shown qualitatively in 
Fig. \ref{rhocha}b). 

The proton structure derived from the REGPEP approach was used to contruct a  
model successfully applied to describe several aspects of proton-proton 
collisions with high multiplicities \cite{kubi14,kubi15,gla16} and also inspired 
the model presented in \cite{deb20} and used here. In the next section we will 
discuss this model in more detail.

\section{The Glauber model for proton-proton collisions}

In this section we briefly describe the model proposed in \cite{kubi15,kubi14} 
and successfully applied in \cite{deb20}.  It combines the parton spatial 
distribution in the proton advanced in \cite{kubi15,kubi14} with the 
geometry machinery provided by the 
Glauber model \cite{glau07}. The quarks are anisotropically distributed, at the 
edges of the Y shape junction  and the gluons are isotropically  
distributed arround the center \cite{kubi15,kubi14}.

Originally, the Glauber model was designed to represent the geometry of 
heavy-ions,  including a dynamical component given by the nucleon cross 
sections. In \cite{loi16} it was adapted to proton-proton collisions.  
The effective number of partonic (subnucleonic) degrees of freedom was 
called $N_c$. The analysis of data performed in 
\cite{loi16}  indicated that this number is $ N_c  \simeq 3 - 10$. 
Following \cite{loi16} and \cite{deb20} 
we will work with the effective number of partonic degrees of freedom
(here we follow \cite{deb20} and call it $N_g$)  and  
reinterpret $N_{part}$ and $N_{coll}$ as  ``number of partons that 
participate in the collision'' and ``number of binary collisions between 
partons'' respectively.

\begin{figure}
\begin{tabular}{ccc}
    \includegraphics[scale=0.68]{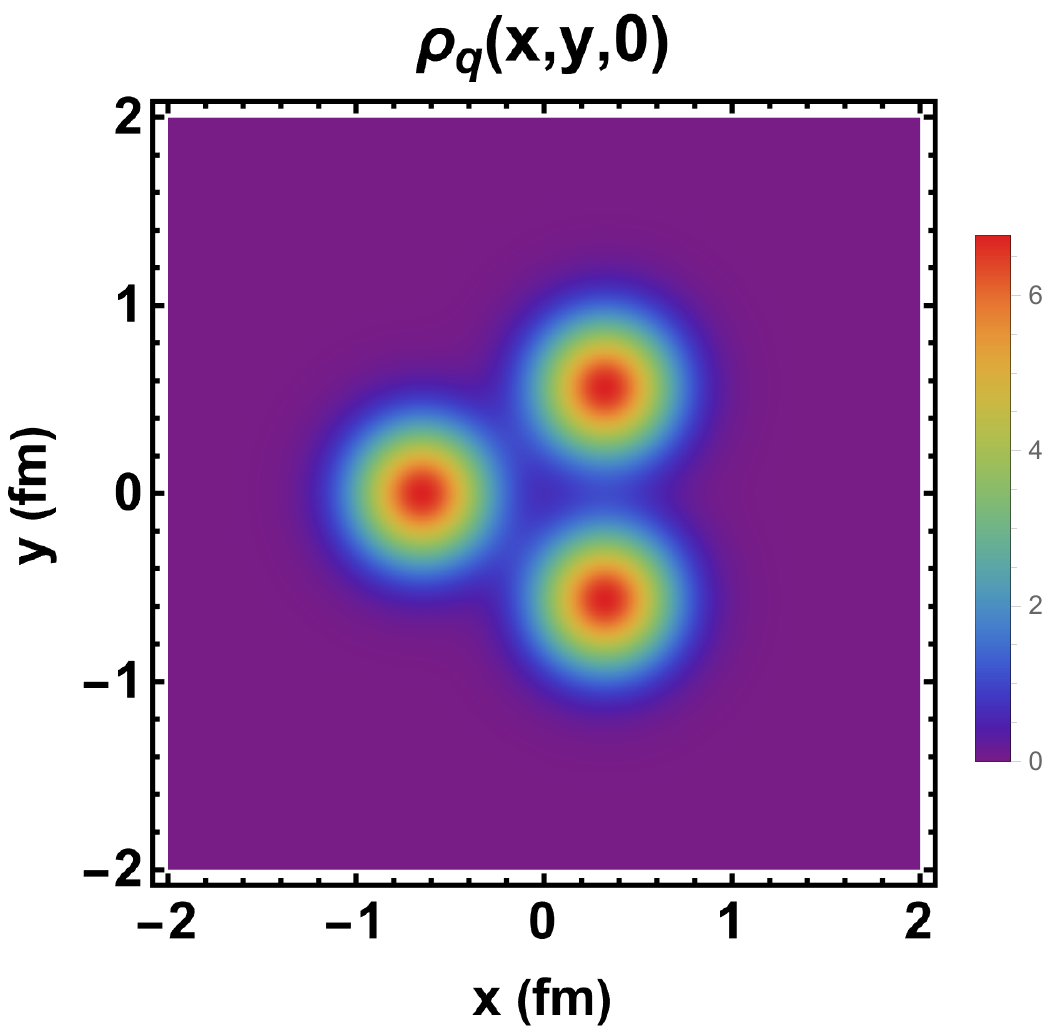}& 

                                              & 
    \includegraphics[scale=0.68]{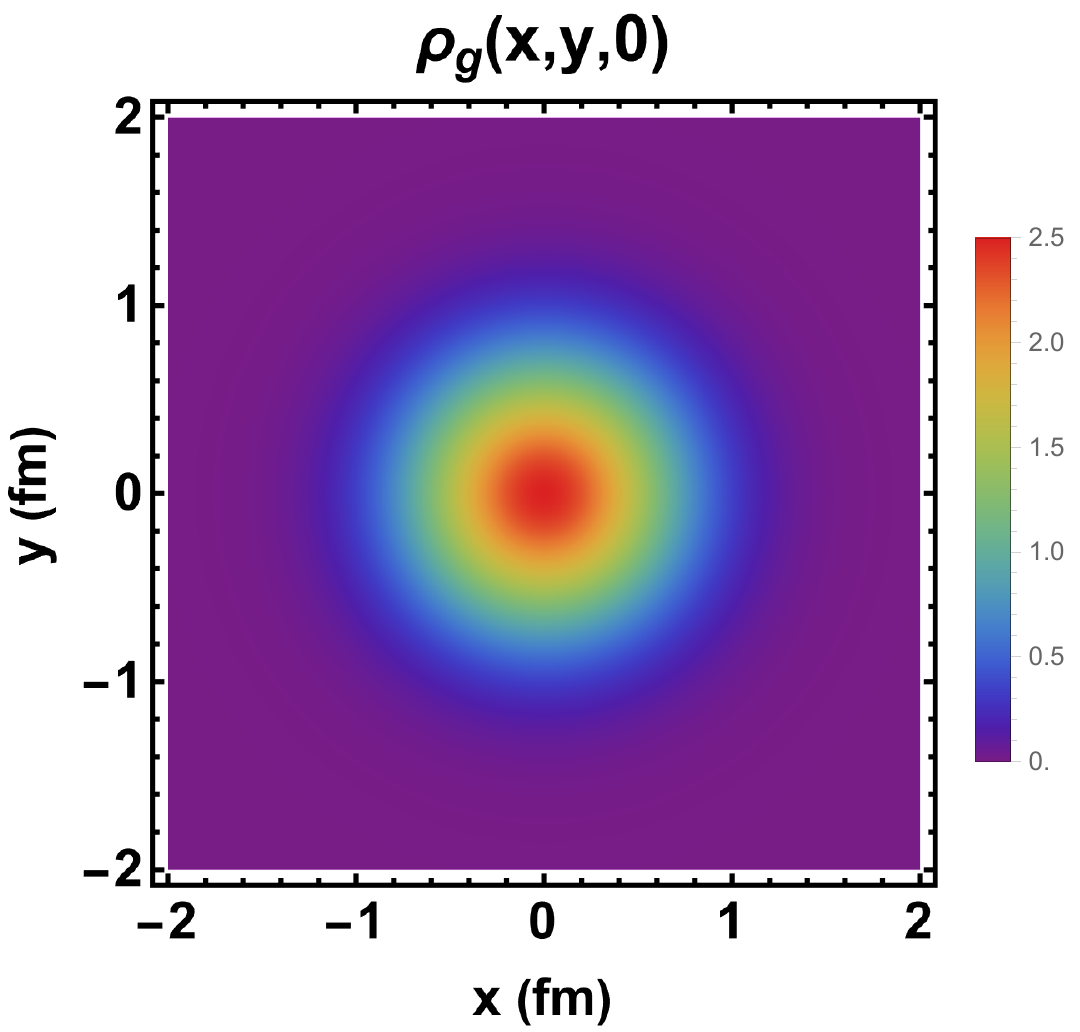} \\
  (a) & \,\,\, & (b)
\end{tabular}
    \caption{a) Quark  and b) gluon  distributions in the transverse plane.}
\label{rhos1}
\end{figure}

\begin{figure}[!ht]
        \centering
\includegraphics[scale=0.68]{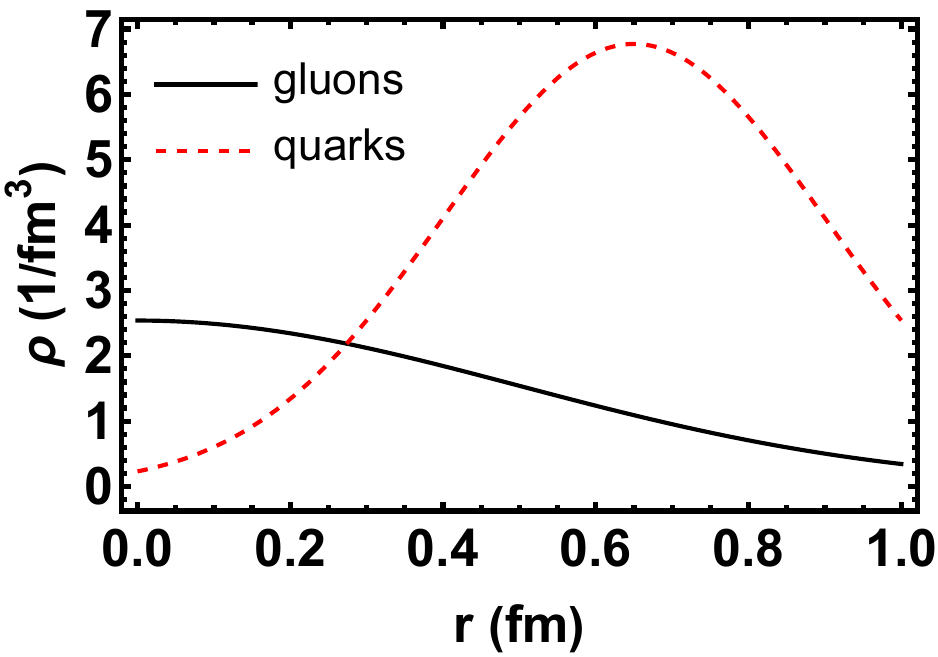}
        \caption{Quark and gluon distributions. Projection of Fig.~\ref{rhos1}
along a diagonal direction, starting from the center of the proton and 
passing through  one maximum of the quark density.}
        \label{rhos2}
\end{figure}

In order to represent the proton internal structure, the matter distribution  
is given by \cite{kubi14}:
\begin{equation} 
\rho_p(\mathbf{r};\mathbf{r}_1,\mathbf{r}_2,\mathbf{r}_3)= \sum_{i=1}^3  
\rho_q(\mathbf{r}-\mathbf{r}_i)+\rho_g \bigg(\mathbf{r}-\sum_{i=1}^3 
\frac{\mathbf{r}_i}{3} \bigg)
	\label{rhop}
\end{equation}
In this distribution, there are three effective quarks, called $i=1,2,3$, 
with the gaussian distribution:
\begin{equation}
	\rho_q(r)=(1-\kappa)\frac{N_g}{3}
\frac{e^{-r^2/2r_q^2}}{(2\pi)^{3/2}r_q^3}
	\label{rhoq}
\end{equation} 
and a gluon gaussian distribution: 
\begin{equation}
	\rho_g(r)=\kappa N_g \frac{e^{-r^2/2r_g^2}}{(2\pi)^{3/2}r_g^3}
	\label{rhog}
\end{equation} 
centered at the average coordinate of the quarks. In these distributions, 
$N_g$ is the total number of partons in the proton, $\kappa$ is the fraction 
of $N_g$ that corresponds to the gluon body, $r_q$ is the radius of the 
effective quark and $r_g$ the radius of the gluon body. These parameters are 
fixed and we take them from previous works with this model. 
In Fig.~\ref{rhos1} we show the contour plot of these distributions.  
Fig.~\ref{rhos2} shows a projection of these distributions and it represents
what we would see moving (up and to the right) from the center of the 
proton to the peak of the quark density (in Fig.~\ref{rhos1}a) plotted 
together with the gluon density from Fig.~\ref{rhos1}b.  It is an 
illustration of the ``core-corona'' aspect of the model and helps to 
understand the results. 

\begin{figure}[!ht]
\begin{tabular}{ccc}
    \includegraphics[scale=0.25]{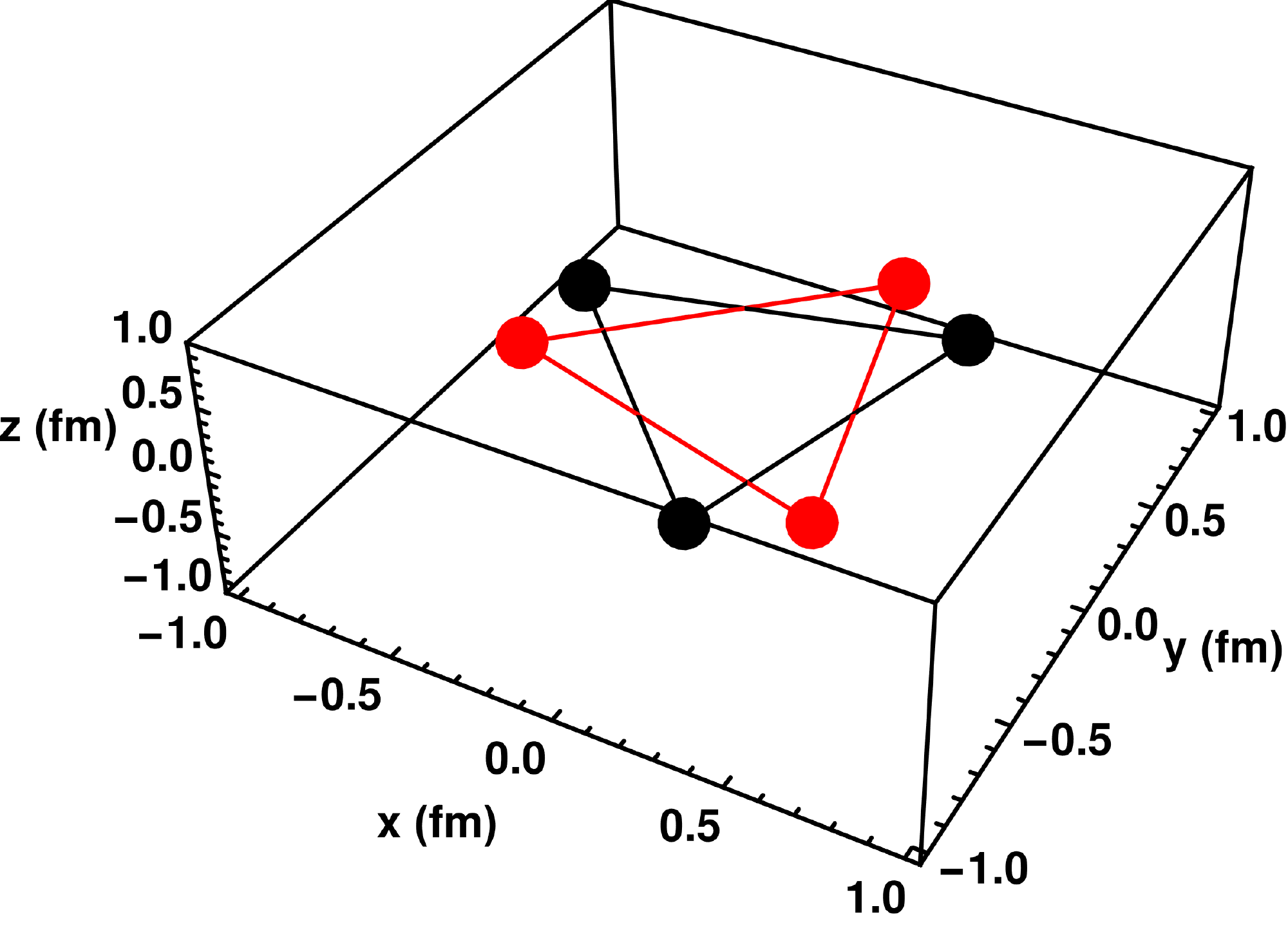}& 
\,\,\,\,\,\,\,\,\,\,\,\,\,\,\,\,\,\,\,\,\,\,\,\, 
\,\,\,\,\,\,\,\,\,\,\,\,\,\,\,\,\,\,\,\,\,\,\,\,
                                              & 
    \includegraphics[scale=0.6]{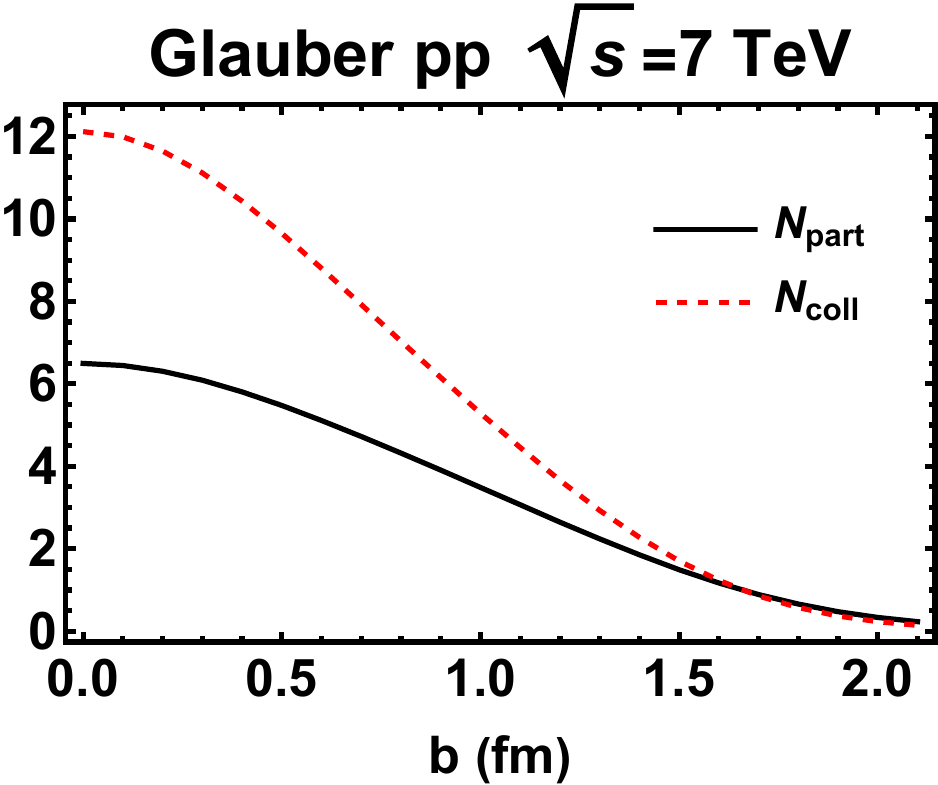} \\
  (a) & \,\,\, & (b)
\end{tabular}
    \caption{a) Example of initial spatial configuration of the two 
colliding protons. b) Average  of $N_{part}$ and $N_{coll}$ over different
quark spatial  configurations for each impact parameter.}
\label{cc-glau}
\end{figure}

With the quark and gluon densities we can calculate the inputs to be used
in the Glauber model. The proton thickness, $T_p(x,y)$, is given by:
\begin{equation}
	T_p (x,y)= \int \rho_p (x,y,z) dz =T_p^q (x,y)+ T_p^g (x,y)
	\label{Tp}
\end{equation} 
where $T_p^q (x,y)$ and $T_p^g (x,y)$ come from the quark and gluon terms 
of Eq.(\ref{rhop}) and are given by: 
\begin{equation}
T_p^q(x,y)= \frac{N_g(1-\kappa)}{6\pi r_q^2} \sum_{i=1}^3 
e^{-\frac{(x-x_i)^2+(y-y_i)^2}{2r_q^2}}
	\label{Tpq}
\end{equation}
and
\begin{equation}
T_p^g(x,y)=\frac{N_g\kappa}{2\pi r_g^2} e^{-\frac{(x-\sum_{i=1}^3x_i/3)^2
+(y-\sum_{i=1}^3y_i/3)^2}{2rg^2}}
	\label{Tpg}
\end{equation}
The overlap function can be calculated as:
\begin{equation}
T_{pp\prime}(b)=\int T_p(x-b/2,y)T_{p\prime}(x+b/2,y)dxdy
\label{Tpp}
\end{equation}
The above equation can be split into four terms. Because of the gaussian forms 
involved, all the integrals can be done analytically. The quark-quark term is 
given by: 
\begin{equation}
T_{pp\prime}^{qq}(b)= \int T_p^q(x-b/2,y) T_{p\prime}^q(x+b/2,y) dxdy
= \frac{N_g^2 (1-\kappa)}{36\pi r_q^2}          
\sum_{i=1}^3\sum_{j=1}^3 \exp \bigg(-\frac{\big(b+(x_i-x_j)\big)^2
+\big(y_i-y_j\big)^2}{4 rq^2} \bigg)
\label{Tqq}
\end{equation} 
with the quarks $i=1,2,3$ from the proton $p$, and $j=1,2,3$ from the proton 
$p\prime$. The gluon-gluon term can be calculated as:
\begin{eqnarray}
T_{pp\prime}^{gg}(b) &=& \int T_p^g(x-b/2,y) T_{p\prime}^g(x+b/2,y) dxdy 
 \nonumber \\
                     &=& \frac{N_g^2 \kappa^2 }{4\pi r_g^2} \exp     
\bigg(-\frac{\big(b+\sum_{i=1}^3 x_i/3 -\sum_{j=1}^3 x_j/3\big)^2 
+\big(\sum_{i=1}^3 y_i/3 -\sum_{j=1}^3 y_j/3\big)^2}{4rg^2}\bigg)
\label{Tgg}
\end{eqnarray}
The gluon-quark term, which accounts for the interactions between the 
gluons from $p$ with quarks from $p\prime$, is given by:
\begin{eqnarray}
T_{pp\prime}^{gq}(b) &=& \int T_p^g(x-b/2,y) T_{p\prime}^q(x+b/2,y) dxdy
\nonumber \\
&=& \frac{ N_g^2 \kappa (1-\kappa)}{6\pi (r_q^2+r_g^2)} 
\sum_{j=1}^3 \exp \bigg(-\frac{\big(b+\sum_{i=1}^3 x_i/3 -x_j \big)^2 
+ \big(\sum_{i=1}^3 y_i/3-y_j\big)^2}{2(r_q^2+r_g^2)}\bigg)
\label{Tgq}
\end{eqnarray}
The last term refers to the interaction between the quarks from 
$p$ and the gluons from $p\prime$. It is given by:
\begin{eqnarray}
T_{pp\prime}^{qg}(b) &=& \int T_p^q(x-b/2,y) T_{p\prime}^g(x+b/2,y) dxdy
\nonumber  \\
                     &=& \frac{N_g^2 \kappa (1-\kappa)}{6\pi (r_q^2+r_g^2)}  
\sum_{i=1}^3 \exp \bigg(-\frac{\big(b+x_i-\sum_{j=1}^3 x_j/3   \big)^2 + 
\big(y_i-\sum_{j=1}^3 y_j/3\big)^2}{2(r_q^2+r_g^2)}\bigg)
\label{Tqg}
\end{eqnarray}
The total thickness function of the model is obtained  by inserting  
Eqs.~(\ref{Tqq}), (\ref{Tgg}), (\ref{Tgq}) and (\ref{Tqg}) into 
Eq.(\ref{Tpp}). It reads: 
\begin{equation}
T_{pp\prime}(b)=T_{pp\prime}^{qq}(b)+T_{pp\prime}^{gg}(b)+T_{pp\prime}^{gq}(b)
+ T_{pp\prime}^{qg}(b)
\label{Tppfinal}
\end{equation}
The number of binary collisions is then given by: 
\begin{equation}
N_{coll}=T_{pp\prime}(b)\sigma^{pp}
\label{Ncollp}
\end{equation} 
where $\sigma^{pp}$ is the parton-parton cross section, which is a parameter 
often used in the literature. The obtained values in the wounded quark model of
Ref.~\cite{sigpp1}, in the hydrodynmical analysis of Ref.~\cite{sigpp2} and 
in the parton transport model of Ref.~\cite{sigpp3} are all in the range 
$\sigma^{pp} = 5 - 10$ mb. The number of participants is calculated according to
\begin{equation}
N_{part}(b) =  N_{coll}^x(b)
\label{Npart_p}
\end{equation}
where $x =0.75$ \cite{glau07}.  In the above equations, $N_{part}$ and 
$N_{coll}$ are calculated for a given quark configuration, i.e. a given 
choice of the quark coordinates in the projectile proton 
$\mathbf{r}_i = (x_i, y_i)$ and in the target proton
$\mathbf{r'}_i = ({x'}_i, {y'}_i)$. In order to simulate a 
proton-proton collision we must now take the average over these configurations. 
This is best done writting the quark position in polar coordinates:
\begin{equation}
\mathbf{r}_i=\frac{d}{2}\big(\cos(\phi_i + \alpha), 
\sin(\phi_i + \alpha) \big)
\,\,\,\,\,\,\,\,\,\,\,\,\,
\mbox{and}
\,\,\,\,\,\,\,\,\,\,\,\,\,
\mathbf{r'}_i=\frac{d}{2}\big(\cos(\phi_i + \beta), 
\sin(\phi_i + \beta)\big)
\end{equation}
where $\phi_1 = \pi/3$, $\phi_2 = -\pi/3$ and $\phi_3 = - \pi$.
This choice of positions generates protons in which the quarks are in  
the vertices of equilateral triangles in the $x-y$ plane, rotated by 
the angles $\alpha$ and $\beta$ ($\alpha\text{, }\beta \in  [0,2\pi]$)  
around the $z$ axis. The angles are randomly chosen. For $d=1.3 \text{ fm}$, 
one example of initial condition is shown in Fig. \ref{cc-glau}a. 
Using the parameters of Table \ref{table1} and fixing the impact parameter 
we choose the angles $\alpha$ and $\beta$. With them we calculate the 
coordinates $\mathbf{r}_i$ and then the thickness function,  which we 
substitute into Eqs. (\ref{Ncollp}) and (\ref{Npart_p})  obtaining
$N_{coll}$ and $N_{part}$. We repeat this procedure for 10000 choices 
and take the average. After that, we move to the next impact parameter and 
repeat the steps. In the end we obtain $N_{coll}$ and $N_{part}$ as a function 
of the impact parameter. The results  for $pp$ collisions 
at $\sqrt{s}=7 \text{ TeV}$ are shown in Fig. ~\ref{cc-glau}b.   
Here the impact parameter is such that $b \in  [0,2.2] \text{ fm}$, with steps 
$\Delta b=0.1 \text{ fm}$. The same procedure was repeated for  
$\sqrt{s}=13 \text{ TeV}$. The resulting curves are quite similar to those 
shown in Fig. ~\ref{cc-glau}b but with higher maximum values. For conciseness 
we do not show them here.  
\begin{table}[ht!]
\caption{Input used in the parton densities and in the Glauber model. See text
for definitions.}
\label{table1}
\centering
    \begin{tabular}{| c | c |}
    \hline
    $N_g$ \cite{deb20}           & $10$ \\ \hline
    $\kappa$   \cite{deb20}      &  $0.5$  \\  \hline
    $r_q \text{ (fm)}$\cite{deb20}&  $0.25$  \\ \hline
    $r_g \text{ (fm)}$ \cite{kubi14} &  $0.5$  \\ \hline
$\sigma^{pp}\text{ (mb)}$ ($\sqrt{s}= 7$ TeV) \cite{sigpp1,sigpp2} &  $4.3$    \\ \hline
$\sigma^{pp}\text{ (mb)}$ ($\sqrt{s}= 13$ TeV) \cite{sigpp1,sigpp2} &  $7.6$    \\ \hline
    $d \text{ (fm)}$  \cite{deb20}& 1.3 \\ \hline
    $x$ \cite{glau07} & 0.75 \\ \hline
    \end{tabular}
\end{table}

\section{Charmonium production}
 
\begin{figure}[!ht]
\centering
\includegraphics[scale=0.5]{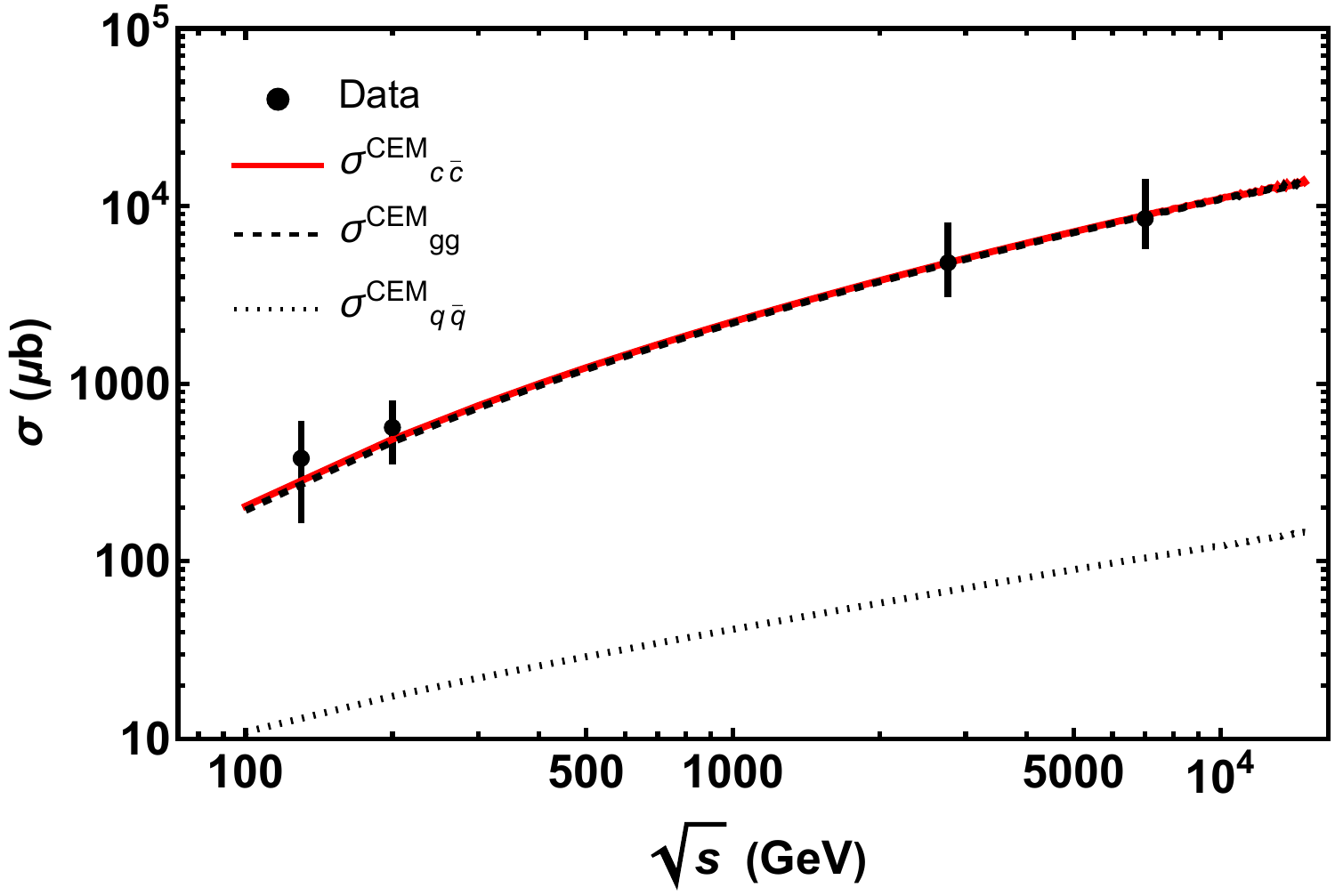}
\caption{Charm production cross sections. Contribution of 
$gg \rightarrow c\bar{c}$ (dashed line),
$q \bar{q} \rightarrow c\bar{c}$ (dotted line) and total cross-section  
(solid line). The experimental data are from 
\cite{phenix02,phenix06,alice12}.}
\label{crosscc}
\end{figure}

Charm production can be described by perturbative QCD and there are currently 
several calculations which reproduce the data with very good accuracy. For 
our purposes a leading order calculation is sufficient. Therefore we shall 
employ the Color Evaporation Model \cite{vogt18,vogt07}  (CEM). 
This model provides a 
way to calculate the production of charm pairs through the processes   
$gg \rightarrow c\bar{c}$ and $q \bar{q} \rightarrow c\bar{c}$.  
The cross section for $J/\psi$ production is then given simply by:

\begin{eqnarray}
\sigma^{CEM} &=& \mathbf{F} K \sum_{i,j}\int_{(2m_c)^2}^{(\Lambda)^2} dm^2 
\int dx_1 dx_2 f_i(x_1,\mu^2) f_j(x_2,\mu^2) \sigma_{ij}(m^2)  
\delta(m^2-x_1 x_2 s)
\nonumber  \\
             &=& \sigma_{gg}^{CEM} + \sigma_{q\bar{q}}^{CEM}
\label{sigcem}
\end{eqnarray} 
where $f(x,\mu^2)$ are the parton distribution functions at the 
renormalization scale $\mu$ and $\Lambda$ is a cut-off. In the case of open
charm production $\Lambda = \sqrt{s} $ and for $J/\psi$ production 
$\Lambda = 2 \, m_D$. The parameter $F$ is equal to the percentage 
of the $c \bar{c}$ states with $ 2 \, m_c < m <  2 \, m_D$ which becomes 
a $J/\psi$. We will assume that $m_c=1200 \text{ MeV}$ and 
$m_D= 1800 \text{ MeV}$.    
The symbol $\sigma_{ii}(m^2)$ represents  the elementary 
$gg \rightarrow c\bar{c}$  and $q \bar{q} \rightarrow c\bar{c}$  
cross sections.  At leading order, they  are given by: 
\begin{equation}
\sigma_{gg}(m^2)=\frac{\pi \alpha_s^2(m^2)}{3m^2} \bigg\{  
\big(1+\frac{4m_c^2}{m^2}+\frac{m_c^4}{m^4} \big) \ln\bigg( 
\frac{1+\lambda}{1-\lambda}\bigg) -\frac{1}{4} \big(7+
\frac{31 m_c^2}{m^2}\big) \lambda \bigg\}
\label{sigmagg}
\end{equation}
and 
\begin{equation}
\sigma_{q\bar{q}}(m^2)=\frac{8\pi \alpha_s^2(m^2)}{27m^2} 
\bigg(1+\frac{2m_c^2}{m^2}\bigg)\lambda
\label{sigmaqq}
\end{equation}
where $\alpha_s$ (the strong coupling constant) and $\lambda$ are 
given by:
\begin{equation}
\alpha_s(\mu^2)=\frac{12 \pi}{(33-2N_f)\ln \big( \frac{\mu^2}
{\Lambda_{QCD}^2}\big)}
\,\,\,\,\,\,\,\,\,\,\,\,\,\,\,\,\,\,\,\,\,\,\,\,\,\,\,\,\,\,\,\,\,\,\,\,\,\,
\lambda =\bigg( 1 - \frac{(2m_c)^2}{m^2} \bigg)^{1/2}
\label{alpha}
\end{equation} 
where $N_f$ is the number of flavors and $\Lambda_{QCD}=200 \text{ MeV}$. 
The parameter $K$ is introduced to account for higher order corrections. 
We can fix it  setting $F=1$, $\Lambda= \sqrt{s}$ and adjusting 
the  cross section Eq.~(\ref{sigcem}) to the experimental data on open charm 
production \cite{phenix02,phenix06,alice12}. 
The result is shown in Fig.~\ref{crosscc}. We obtain a good fit          
of these data with $K=2$. Imposing that $\lambda$ is real leads to the 
kinematical constraint 
$x_1 x_2 s \ge 4m_c^2$. Since the smallest value of $x_1$ occurs for 
$x_2=1$, the parton momentum fraction must be such that 
$(2m_c)^2/s \le x_1 \le 1$ and $(2m_c)^2/(sx_1) \le x_2 \le 1$. The parton 
distribution functions, used with $Q^2=2.4^2 \text{ GeV}^2$, are taken from 
the CTEQ5 set \cite{PDFs}. Finally, the number of $c \bar{c}$ pairs is given by: 
\begin{equation}
N_{c\bar{c}}(b)=T_{pp\prime}^{gg}(b) \,  \sigma_{gg}^{CEM} + 
T_{pp\prime}^{qq}(b) \,  \sigma_{qq}^{CEM}
\label{Ncc}
\end{equation}
The $T_{pp\prime}^{qg}(b)$ and $T_{pp\prime}^{gq}(b)$ terms of            
Eq.(\ref{Tppfinal}) are not included because the quark-gluon interactions 
do not produce $c\bar{c}$ pairs. 

\section{Results} 

We now address the experimental data of \cite{alice-psi7,alice-psi13,alice-cc}. 
Following the experimental papers, we present the yields  
 in terms of the normalized pseudo-rapidity and rapidity 
densities:
\begin{equation}
\frac{({dN}/{d\eta}(\eta=0))}{\langle {dN}/{d\eta}  \rangle}
\,\,\,\,\,\,\,\,\,\,\,\,\,\,\,\,\,\,\,\,\,\,\,\,\,\,\,\,\,\,\,\,\,\,\,\,
\frac{({dN}/{dy}(y=0))}{\langle {dN}/{dy}  \rangle}
\label{rapdens}
\end{equation}
The charged particle pseudorapidity density at $\eta=0$ is calculated as in 
\cite{kn01}:
\begin{equation}
\frac{dN}{d\eta} (\eta=0) = 
n_{pp}(s) \big\{(1-f)\frac{N_{part}}{2} + fN_{coll}\big\}
\label{dndeta}
\end{equation} 
where $N_{part}$ and $N_{coll}$ are given by  (\ref{Npart_p}) and (\ref{Ncollp})
respectively and $n_{pp}(s)$ is  given by \cite{kn01}:
\begin{equation}
n_{pp}(s)=2.5-0.25 \log[s] + 0.023 (\log[s])^2
\label{npp}
\end{equation}
The factor $f$ is the fraction of hard processes in the collision and can be 
estimated through the ratio: $f = \sigma_{minijet} / \sigma_{inel}$ as in 
\cite{mini1,mini2}. A minijet is defined as the result of a parton-parton 
collision with $p_T > p_{T0}$ with $p_{T0}$  being of the order of a few 
GeV (with a possible dependence on $\sqrt{s}$). 
The average density $ \langle dN/d\eta  \rangle$ is a number  
given by each experimental group and we show them in Table II, together with
the $F$ parameter.
\begin{table}[ht!]
\caption{Input used for $J/\psi$ production. See text for definitions.}
\label{table2}
\centering
    \begin{tabular}{| c | c | c |}
    \hline
           & $\sqrt{s} = 7$ TeV &    $\sqrt{s} = 13$ TeV                \\ \hline
        f  & 19 \%  \cite{mini1,mini2} &   16 \%  \cite{mini1,mini2}    \\ \hline
        F  & 1.5 \% \cite{vogt18,vogt07}  & 4.3 \% \cite{vogt18,vogt07} \\ \hline
$ \langle dN /d \eta \rangle$ & $6.1$ \cite{alice-dn-13} 
& $6.4$ \cite{alice-dn-13} \\ \hline
$ \langle dN_{J/\psi}/dy \rangle$ & $8.2 \times 10^{-5}$ \cite{alice-psi7} 
& $7.9 \times 10^{-5}$ \cite{alice-psi13}
\\ \hline
    \end{tabular}
\end{table}

\begin{figure}[h]
\begin{tabular}{ccc}
    \includegraphics[scale=0.53]{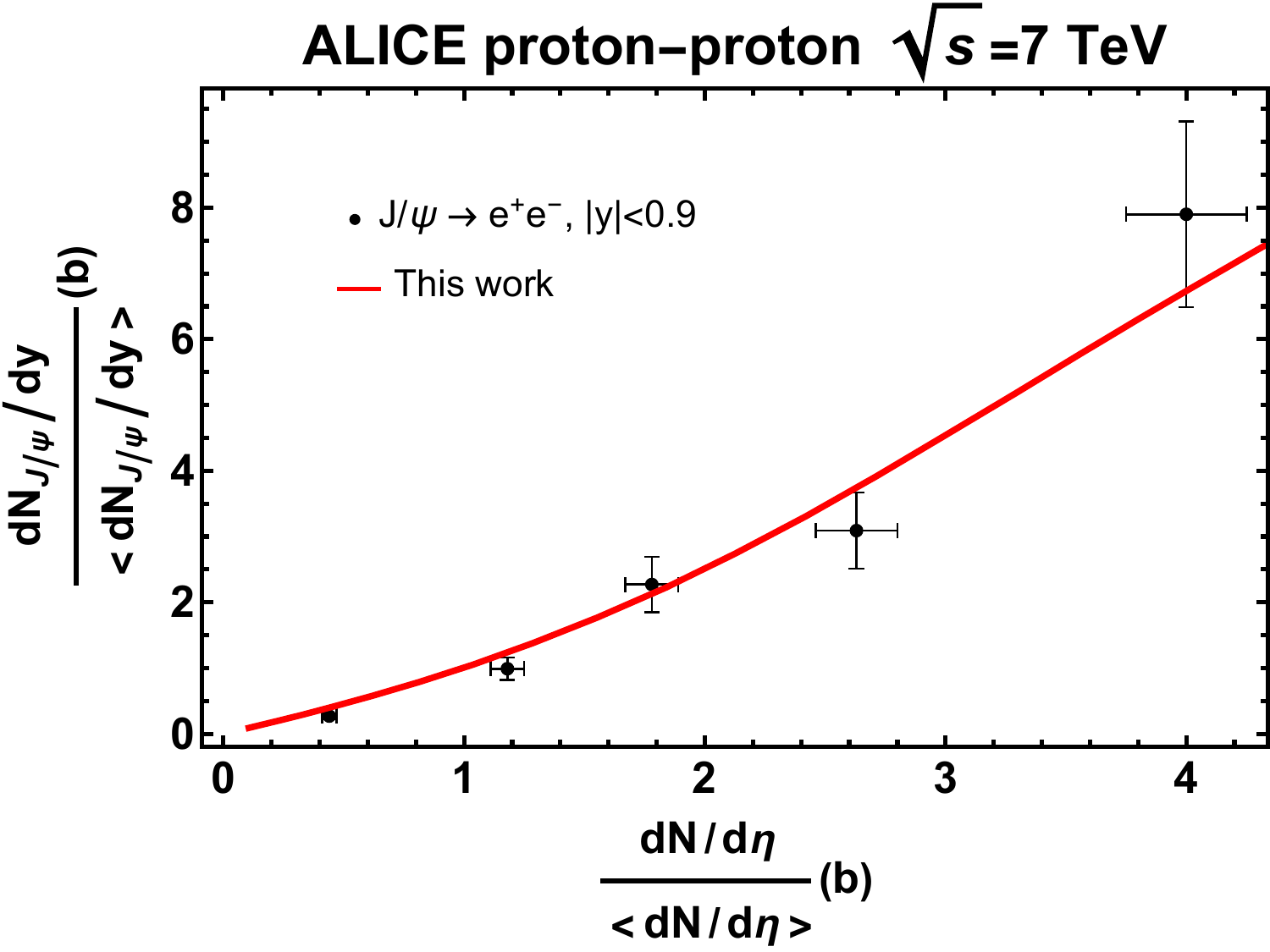}& 
                                             & 
    \includegraphics[scale=0.53]{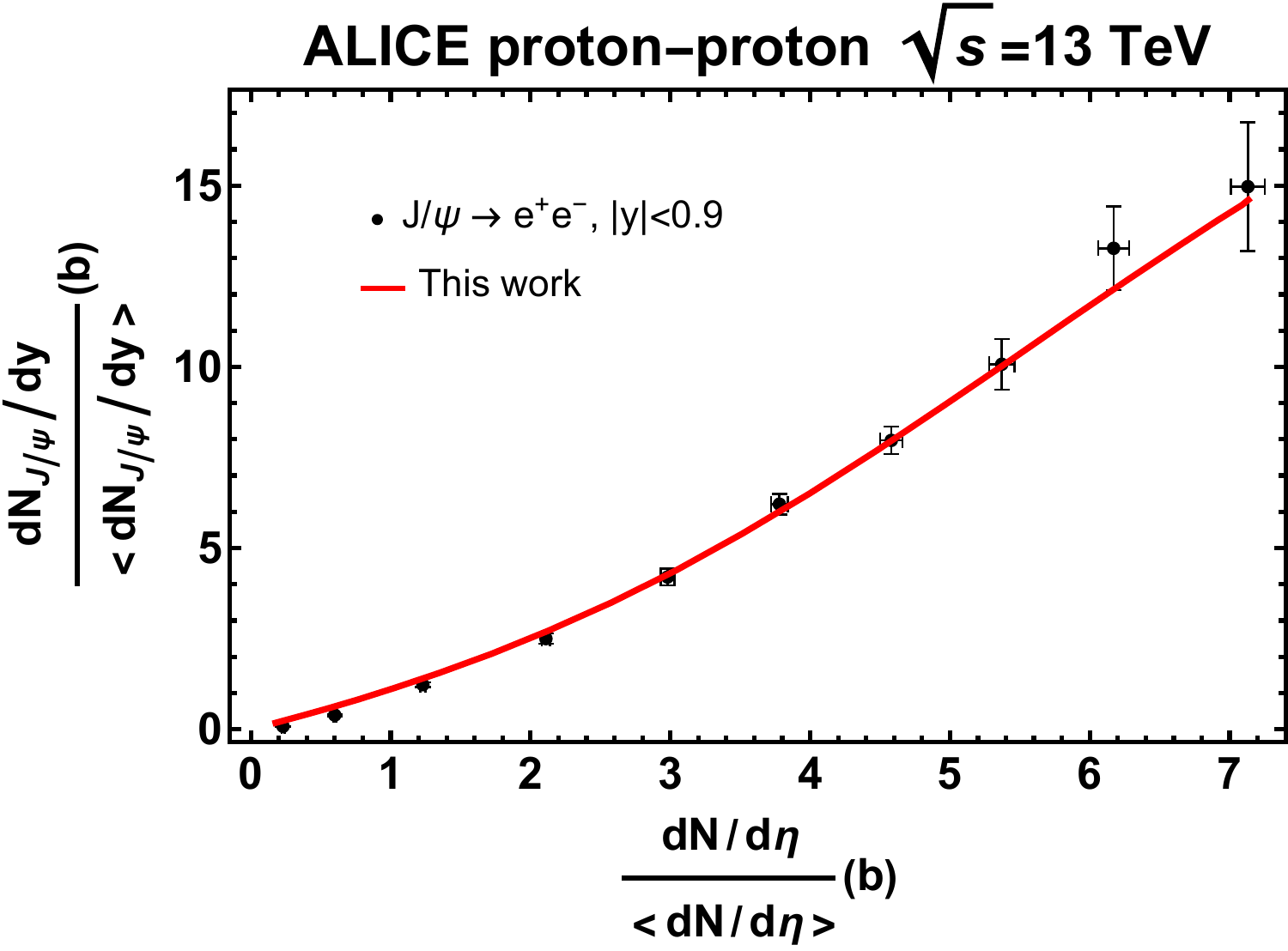} \\
  (a) & \,\,\, & (b)
\end{tabular}
\caption{
a) $J/\psi$ relative yield in $pp$ collisions at $\sqrt{s}_{NN}=7 \text{ TeV}$. 
The experimental data are from \cite{alice-psi7}. 
b) $J/\psi$ relative yield in $pp$ collisions at $\sqrt{s}_{NN}=13 \text{ TeV}$. 
The experimental data are from \cite{alice-psi13}.} 
\label{jpsiexp}
\end{figure}

The $J/\psi$ rapidity density, $dN_{J/\psi}/dy$, is obtained from the differential 
form of  Eq.(\ref{Ncc}):
\begin{equation}
    \frac{dN_{c\bar{c}}}{dy}(b)=T_{pp\prime}^{gg}(b) \,    
\frac{d \sigma_{gg}^{CEM}}{dy} + T_{pp\prime}^{qq}(b) \,  
\frac{d\sigma_{qq}^{CEM}}{dy}
    \label{dNdy}
\end{equation}
We can evaluate this expression starting from Eq.(\ref{sigcem}) and applying the
following change of variables:
\begin{equation}
x_1 = \frac{p_T}{\sqrt{s}} e^y 
\,\,\,\,\,\,\,\,\,\,\,\,
x_2 = \frac{p_T}{\sqrt{s}} e^{-y}
\,\,\,\,\,\,\,\,\,\,\,\,
x_1 x_2 s = p_T^2
\,\,\,\,\,\,\,\,\,\,\,\,
dx_1 dx_2 \rightarrow \frac{2p_T}{s} dy dp_T 
\label{variables}
\end{equation}
After changing the variables from $(x_1,x_2)$ to $(y,p_T)$, 
we integrate over $p_T$, differentiate with respect to $y$ and take $y=0$.
The results for the $J/\psi$ yields for $\sqrt{s}=7 \text{ TeV}$ and 
$\sqrt{s}=13 \text{ TeV}$ are shown in Fig.~\ref{jpsiexp}.  As it can be seen
we obtain a very good description of data. One might argue that we have too 
many input parameters and this reduces the predictive power of the model. 
However, all these input numbers are strongly constrained by other independent
studies of other observables. As we notice in the tables, all the numbers are 
consistent with the equivalent numbers found elesewhere. We took the precaution 
of nowhere using exotic values for any of these numbers.  In view of the results, 
we conclude that the $J/\psi$ dependence on the charged multiplicity is 
compatible with the geometrical picture of the proton used here.  
The analysis can be
extended to $D$ production, to beauty production and to different cuts in $p_T$
and rapidity. We are already working in these topics. 

\section{Concluding remarks}

We have developed the idea that proton-proton collisions can be described  
by Y-Y collisions. Averaging over orientations this yields a circular 
configuration with an inner gluonic shell and an outer quark shell.  This is  
a core (gluon) - corona (quark) model of the proton. Going from peripheral   
to more central and then to ultra-central, we go from quark-quark collisions 
to gluon-gluon collisions. Since gluons are much more abundant, and since 
$ \sigma (gg \to c \bar{c}) \, >> \, \sigma (q \bar{q} \to c \bar{c})$
the cross sections grow strongly. These effects combined  explain the
growth seen in the data.

This behavior was qualitatively expected and we have used a combination of 
models to implement this idea quantitatively: a model for the parton spatial 
densities \cite{deb20}; the Glauber model for proton-proton collisions  
\cite{loi16} and the color evaporation model for charmonium production 
\cite{vogt18,vogt07}. These models had been previously tested
in other contexts and were shown to give good results. All the parameters 
used were strongly constrained by the analysis of other data and here we 
had little room to change them.  We obtained a good description of 
charmonium production in proton-proton events with high multiplicities. 
This encourages us to further extend this model to open charm production 
and to proton-nucleus collisions.

\begin{acknowledgments}
We are grateful to K. Werner for instructive discussions. 
This work was financed by the Brazilian funding agencies CNPq, FAPESP and 
by the INCT-FNA.
\end{acknowledgments}


\end{document}